\documentclass[12pt,a4paper]{article}

\usepackage[british]{babel}

\usepackage[a4paper,top=2cm,bottom=2cm,left=2.5cm,right=2.5cm,marginparwidth=1.75cm]{geometry}

\usepackage[style=apa, backend=biber]{biblatex} 
\DeclareLanguageMapping{british}{british-apa} 
\DeclareFieldFormat[article]{volume}{\apanum{#1}} 

\usepackage{amsmath}
\usepackage{graphicx}
\usepackage[colorlinks=true, allcolors=blue]{hyperref}
\usepackage{hyperref}
\usepackage[title]{appendix}
\usepackage{mathrsfs}
\usepackage{amsfonts}
\usepackage{booktabs} 
\usepackage{caption}  
\usepackage{threeparttable} 
\usepackage{algorithm}
\usepackage{algorithmicx}
\usepackage{algpseudocode}
\usepackage{listings}
\usepackage{enumitem}
\usepackage{chngcntr}
\usepackage{booktabs}
\usepackage{lipsum}
\usepackage{subcaption}
\usepackage{authblk}
\usepackage[T1]{fontenc}    
\usepackage{csquotes}       
\usepackage{diagbox}
\usepackage{comment}

\usepackage{setspace}
\usepackage{titlesec}
\titleformat{\section} 
  {\normalfont\Large\bfseries}{\thesection.}{1em}{}
  
\usepackage{lineno} 

\rightlinenumbers 

\linenumbers 

\usepackage{float}   
\usepackage{caption} 


\title{Asymptotic Behavior in High School Physics: physics insights and discussions}

\author[1]{Kyle Kou Yuchang}
\author[2,*]{Paul Zhang Yixing}
\author[2]{Victor Wang Shuang}
\affil[1]{\small Wuhan Britain-China Senior High School, No.10 Gutian Ce Rd., Qiaokou District, Wuhan, Hubei, P.R.China}
\affil[2]{\small Physics Department, Wuhan Britain-China Senior High School, No.10 Gutian Ce Rd., Qiaokou District, Wuhan, Hubei, P.R.China}

\date{13 Febuary 2025}  

\begin{document}
\nolinenumbers
\maketitle

\begin{abstract}
Asymptotic analysis provides powerful insights into physical systems by examining their behavior in limiting cases. This paper explores how extending this advanced methodology to high school physics education can deepen conceptual understanding of fundamental topics. Through two carefully selected case studies - multi-ball collisions and internal resistance in circuits - we demonstrate how asymptotic approaches offer:
\begin{itemize}
    \item Intuitive physical interpretations beyond standard derivations
    \item Resolution of conceptual paradoxes through limit analysis
    \item Connections between elementary and advanced physics concepts
\end{itemize}
Our analysis reveals that asymptotic methods help students develop stronger physical intuition while preparing them for more advanced studies. By examining boundary behaviors in collision dynamics and circuit theory, we show how these techniques transform abstract equations into tangible physical understanding, suggesting valuable applications across the high school physics curriculum.
\end{abstract}


\section*{Nomenclature}

\begin{tabbing}
$u_i$ \qquad \= Initial velocity of ith object\\
$v_i$ \qquad \=  Final velocity of ith object\\
$m_i$  \qquad \=Mass of ith object \\
$\mathcal{E}$\qquad \= Electromotive Force (E.M.F)\\
$ r$ \> Internal resistance of the cell\\
$ R $ \> External resistance in the loop \\
$I$ \> Current in the loop
\end{tabbing}

\section{Introduction}
Asymptotic methods serve as fundamental analytical tools in physics, enabling the investigation of system behavior under extreme parametric limits. These approaches examine how physical quantities evolve as key parameters—such as time, spatial dimensions, or other variables—approach critical values like zero or infinity. By focusing on these limiting cases, asymptotic analysis reveals fundamental principles governing physical phenomena.

Momentum conservation represents a core concept in high school physics curricula. Defined as the product of mass and velocity ($\vec{p} = m\vec{v}$), this vector quantity characterizes an object's motion while governing interaction dynamics. Similarly, internal resistance—the inherent opposition to current flow within power sources—manifests through voltage drops that distinguish real power sources from ideal models.

When mass ratios become extreme in collision scenarios, asymptotic behavior emerges: light objects approach maximum velocity limits (up to twice the heavy object's initial speed). This exemplifies how asymptotic analysis simplifies complex systems by identifying dominant behaviors without solving complete equations. In electrical systems, asymptotic limits clarify battery behavior: as current approaches zero, terminal voltage converges to EMF ($V \to \mathcal{E}$); under high current loads, voltage drop across internal resistance becomes substantial ($V \to 0$). 

These approaches provide pedagogical advantages by:
\begin{itemize}
    \item Developing intuitive understanding of physical limits
    \item Connecting elementary and advanced physical concepts
    \item Resolving apparent paradoxes in boundary cases
\end{itemize}
This paper demonstrates how asymptotic methods enhance conceptual mastery in high school physics while preparing students for advanced studies.

\section{Applying in momentum}\label{sec2}

\subsection{General Problem statement}
The centres of the spheres 1, 2 and 3 lie on a single straight line. Sphere 1 is moving with velocity $u_1$ directed along this line and hits sphere 2. Sphere 2, according after collision a velocity, hits sphere 3. Both collisions are elastic. Find the mass of the sphere 2 for the sphere 3 to acquire maximum velocity (the masses of $m_1$ and $m_3$ of spheres 1 and 3 are known.)

\begin{figure}[!ht]
\centering
\includegraphics[width=0.4\linewidth]{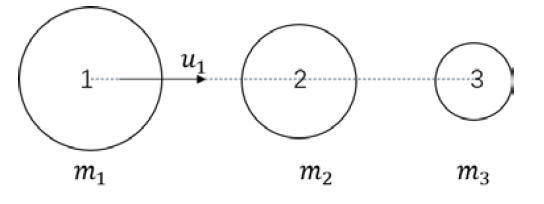}
\caption{\label{fig:3ball}The picture of 3 ball collision problem.}
\end{figure}

\subsection{Possible solutions}\label{subsec1}
Usually the solution shall consider the idea of energy and momentum conservation and get equations from there. After the conservation equations have been written out, the one usually need to find the derivative of the variable with respective to the independent variable, in this case, is the intermediate mass $m_2$.

\subsubsection{Standard solutions}\label{subsubsec1}
Following our terminologies, we have the conservation of momentum in two collsions.
\begin{equation}
    m_1 u_1 = m_1v_1 + m_2 u_2
\end{equation}
Here $u_2$ refers to the velocity of the 2nd ball after first collision. By relative speed approach, we have $u_1 = u_2 -v_1$. 
Hence we have 
\begin{equation}
u_2 = \frac{2m_1u_1}{m_1 + m_2}
\end{equation}
For the collision between the ball 2 and ball 3, we have
\begin{equation}
    u_3 = \frac{2m_2u_2}{m_2 +m_3} = \frac{4 m_1 m_2 u_1}{(m_1 + m_2)(m_2 + m_3)}
    \label{eq3}
\end{equation}. 
By finding the derivative of $u_3$ with respective to $m_2$, we have the following 
\begin{equation}
    \frac{du_3}{dm_2} = \frac{4m_1 u_1 (m_1 + m_2)(m_2 + m_3) - 4m_1 m_2 u_1(m_1 + 2m_2 + m_3)}{[(m_1 + m_2)(m_2 + m_3)]^2}
\end{equation}. \label{masseqn}
Setting the above derivatives to 0, we shall have \begin{equation}
    m_2 = \sqrt{m_1 m_3}
\end{equation}, which is the geometric mean between $m_1$ and $m_3$.

\subsubsection{Intuitive solution: asymptotic behavior} \label{subsubsec2}
The intuitive solution is useful when we are trying to analyzing this question. The one may start to think the velocity of ball 3 when the mass of ball 2 varies. 

There is one solution, which is trivial but not naive, when $m_2 = m_1 = m_3$. It applies that all the energy available has been transferred from ball 1 to ball 3, hence it could be the maximum energy that ball 3 may process. In this case, the ball 3 is sort of the average between the ball 1 and ball 3. 

Next, we shall consider the relationship between $m_2$ and $v_3$. We realize if $m_2 >> m_1$, then $v_3$ shall be close to zero. Similar case happens when $m_2 \xrightarrow{}0$. Therefore we have two ends of $v_3$ to be 0 and $v_3$ is also a continuous function of $m_2$, the maximum should occur in between there.  

Obviously, the maximum does not occur for the arithmetic mean of $m_1$ and $m_3$, as we can consider the situation in which $m_1$ is much greater than $m_3$. In this case, the final velocity for $m_3$ is greater when $m_2 \sim m_1$. Therefore, we shall test the idea of geometric mean, which is the solution for the cases. The idea for geometric mean also arises for the reason that the mass of each object in equation \eqref{masseqn}. It is easier to consider the geometric mean rather than arithmetic mean.

Therefore, we can conclude that the solution to the problem is $ m_2 = \sqrt{m_1 m_3}$. 

The one may also consider harmonic mean as one of the possible candidate. However, harmonic mean shall not be the right answer as it shall be affected by the extreme values. For example, if mHarmonic mean of a list of numbers tends strongly toward the least elements of the collection, it tends (compared to the arithmetic mean) to mitigate the impact of large outliers and aggravate the impact of small ones. In physics, harmonic mean can be used to represent the average speed of the journey when the each trip covers the same amount of distance. In our situation, the harmonic mean may not be a feasible idea, since the constrain for the problem is momentum and energy transferred. These quantities are both directionally proportional to the mass. Unlike average speed situation, the constrain, average speed, is inversely proportional to the time spent, which is the independent variable in the problem. 

\subsubsection{Idea of impedance matching}\label{subsubsec3}
As discussed in  \ref{subsubsec2}, the key part to understand the question is the kinetic energy transferred in each collision. Hence, the one may recall the idea in acoustics and think the idea about impedance matching as discussed in \textcite{Santos2012}. 
We shall give a basic summary about the idea discussed in the paper.  \textcite{Santos2012} used idea of Transmission coefficient and Reflection Coefficient in waves mechanics, and express the amount of energy being transferred out in each collisions. In the end, we shall find out that the amount of energy transferred is given by \begin{equation}
    T_{13} = \frac{16\mu_{13}}{(1 + \mu_{13} + \frac{\mu_{13}}{\mu_{23}}+\mu{23})^2}
\end{equation} \ \label{TransferEqn}.
Referring to the Fresnel coefficients in modern optics,  the one is able to write the above coefficients into the ratio of their optical impedance, \begin{equation}
    t_{13} = \frac{2Z_1}{(Z_3 + Z_1)} = \frac{16Z_{1}Z^{2}_2Z_3}{(Z_1+Z_2)^2(Z_2+Z_3)^2}.
\end{equation}\label{Fresnel}
The maximum transmission occurs when $Z_2 = \sqrt{Z_1Z_3}$.\\
Comparing the \eqref{TransferEqn} and \eqref{masseqn}, the one may find some similarities there. This is because the idea for impedance matching taking places, the one may define the mechanical impedance, which is the response of the ball to the impulse $\Vec{J} = \int_{t_f}^{t_f}dt\Vec{F(t)} = \Vec{P_f} - \Vec{P_i} = m\Vec{v_f}$ , is defined as \begin{equation}
    Z_{mechanical} = \frac{J}{v_f}
\end{equation} \label{mechanicalImpedance}. The one may need to note that it is very important for the target ball to be at rest in the beginning, otherwise the mechanical impedance obtained would not be the form as stated above.

Referring to the idea of impedance matching,  we shall consider the amount of momentum transferred during collision process. The idea for such consideration is by comparing the following equations:
\begin{equation}
    \Phi = L I
\end{equation} and
\begin{equation}
    p = mv
\end{equation}

We shall realize that the term $m$ in the mechanical system is identical to the $L$ term in the electrical system. All of these terms describe the reluctance to change. 

The reason in our consideration for the flux $\Phi$ instead of the resistance $R$ is our mechanical system is on the frictionless surface. Therefore, there arere no resistive forces acting on the ball. The only contribution to the impedance term $Z$ is something similar to the inductance $L$. If we accept the above argument, the one could easily conclude that the optimum transfer occurs when $m_2 = \sqrt{m_1m_3}$.

\subsubsection{Dimension analysis}
The above problem could also be solved by dimension analysis as demonstrated in \cite{Zee}. 

The quantities involved are $m_1$, $m_2$, $m_3$, input to the system is $v_1$ and output to the system is $v_3$. The dimension involved are $M$ and $LT^{-1}$. We can construct the dimensionless quantities $\pi = \frac{v_1}{v_3}$ and $\mu = \frac{m_2}{\sqrt{m_1m_3}}$. From the $\Pi$ theorem, we know that $f(\mu) = \pi$. 

Then we can follow the discussion in intuitive solution, we shall realize when $\mu = 1$, which is the case $m_2 = \sqrt{m_1m_3}$, there is the extreme value for $v_3$, which is the maximum possible velocity.

\section{Asymptotic Behavior in Physical Systems}
\subsection{Continuity Principles}
Asymptotic solutions exhibit continuity akin to perturbation responses: infinitesimal mass variations ($\delta m_2$) produce commensurate velocity changes ($\delta u_3$). This follows from momentum conservation, which prohibits discontinuities in physical observables. Figure \ref{fig:internal} illustrates the smooth $u_3(m_2)$ relationship with well-defined extrema.

\subsection{Boundary Condition Analysis}
The apparent paradox at $m_2 \to 0$ merits examination. While direct collision ($m_1 \to m_3$) yields $u_3 = 2m_1u_1/(m_1+m_3)$, our three-body system requires:
\begin{itemize}
    \item $m_2$ must physically exist ($m_2 > 0$)
    \item Collisions occur sequentially, not concurrently
\end{itemize}
Thus $\lim_{m_2 \to 0} u_3 = 0$ correctly describes the limit where Sphere 2 vanishes, distinct from direct collision scenarios.

\subsection{Internal Resistance Case Study}
The circuit equation $V = \mathcal{E} - Ir$ presents a didactic example of asymptotic behavior. The apparent contradiction at $I=0$ (where Ohm's law suggests $0 = \mathcal{E}$) resolves through asymptotic analysis:
\begin{itemize}
    \item As $R \to \infty$, $I \to 0$ and $V \to \mathcal{E}$
    \item The indeterminate form $0 \cdot \infty = \mathcal{E}$ reflects physical reality
\end{itemize}
Table 1 demonstrates this asymptotic approach numerically, showing terminal voltage converging to EMF as load resistance increases.

\begin{table}[ht]
\centering
\begin{tabular}{ c c c c}
 R & $\mathcal{E}$ & I & V \\ 
 $1\Omega$ & 3.0V & 1.5A & 1.5V\\  
 $2\Omega$ & 3.0V & 1.0A &  2.0V\\ 
 $10\Omega$ & 3.0V & 0.27A &  2.7V\\   
 $20\Omega$ & 3.0V & 0.14A &  2.9V\\   
 $200\Omega$ & 3.0V & 0.015A &  3.0V\\   
 $1000\Omega$ & 3.0V & $3.0\times 10^-3$A &  3.0V\\   
 $10000\Omega$ & 3.0V & $3.0\times 10^-4$A &  3.0V\\   
 $100000\Omega$ & 3.0V & $3.0\times 10^-5$A &  3.0V\\   
\end{tabular}
\caption{\label{table}The table of internal resistance.}
\end{table}

It is quite self-evident that as the current $I\xrightarrow[]{}0$, the voltage on the load resistor is approaching the e.m.f of the cell.

\section{Conclusion}

The exploration of asymptotic behavior in high school physics reveals its profound educational value for developing intuitive understanding of physical systems. Through our analysis of two canonical problems - multi-ball collisions and internal resistance circuits - we demonstrate how asymptotic analysis provides unique insights beyond standard derivations.

In the collision system, we observe that the maximum velocity transfer occurs at the geometric mean ($m_2 = \sqrt{m_1m_3}$), a result elegantly explained through three complementary approaches: 
\begin{enumerate}[label=(\roman*)]
    \item The limiting behavior when mass ratios approach extremes
    \item Mechanical impedance matching analogies
    \item Dimensionless parameter analysis
\end{enumerate}
Each perspective reinforces how asymptotic considerations illuminate the underlying physical principles while avoiding complex calculus.

Similarly, for internal resistance, the terminal voltage's asymptotic approach to EMF ($\mathcal{E}$) as $R \to \infty$ resolves apparent paradoxes in Ohm's law. The tabulated experimental data vividly demonstrates how systems evolve toward limiting cases, transforming abstract equations into tangible physical behavior.

These case studies highlight how asymptotic methods serve as powerful pedagogical tools that:
\begin{itemize}
    \item Reveal fundamental system behaviors obscured in exact solutions
    \item Provide conceptual bridges to advanced topics (impedance matching, perturbation theory)
    \item Develop dimensional analysis skills
    \item Resolve mathematical paradoxes through physical reasoning
\end{itemize}

We recommend incorporating asymptotic analysis throughout high school physics curricula, particularly when introducing:
\begin{itemize}
    \item Limits and continuity concepts in mechanics
    \item Circuit theory with non-ideal components
    \item Model simplification techniques
\end{itemize}
Future work could explore asymptotic approaches in thermodynamics (e.g., ideal gas limits) and optics (ray optics limits of wave theory). By emphasizing how systems behave at boundaries, educators can cultivate deeper physical intuition and prepare students for advanced studies where asymptotic methods form the cornerstone of analytical techniques.

\section*{Acknowledgments}
The authors thank Prof. Wang Qinghai and Dr. Alan Laine for the help and guidance on the way of thinking the problem.


\printbibliography
\end{document}